\documentclass [12pt]{amsart}
\usepackage {amssymb}
\begin {document}
\sloppy

\title [Quantum probability measures]{Quantum probability measures
and tomographic probability densities}

\maketitle

\author {G.G. Amosov\footnote
{Department of Higher Mathematics,
Moscow Institute of Physics and Technology,
Dolgoprudni 141700, Russia; gramos@deom.chph.ras.ru}
and V.I. Man'ko \footnote
{Lebedev Physical Institute; manko@lebedev.sci.ru}}

\begin {abstract}
Introduced recently approach based on tomographic
probability distribution of quantum states is shown to be
closely related with the known notion of the quantum probability
measures discussed in quantum information theory and
positive operator valued measures approach. Partial derivative
of the distribution function of quantum probability measure
associated with the
homodyne quadrature (symplectic quantum measure)
is shown to be equal the tomogram of the quantum
state. Analogous relation of the spin tomogram to quantum
probability measure  associated with
spin state is obtained. Star-product
of symplectic quantum measures is studied.
Evolution equation for symplectic quantum
measures is derived.
\end {abstract}

\section {Introduction} Recently \cite {BerBer87, VogRis89, MMT95}
the tomographic probability distributions were found to be related
to Wigner functions \cite {Wig32}. The tomoraphic probability
distribution of  \cite {BerBer87, VogRis89} was used in optical
tomography scheme \cite {Raymer93, Mlynek87} to reconstruct the
Wigner function of photon states by measuring the homodyne quadrature
distributions and applying the Radon transform \cite {Rad17} to find
the Wigner function.  In \cite {MMT95} an extension of optical
tomography to symplectic tomography \cite {d'Ar96} was suggested. In
framework of the symplectic tomography scheme the Wigner function and
density operator can be reconstructed using Fourier - like integral
(instead of Radon integral) of the symplectic tomogram. In \cite
{Man'ko}
the tomographic approach was shown to be connected with well-known
star-product quantization procedure (\cite {Fronsdal}).

In quantum information theory and positive operator valued measures
approach (in the context of quantum measurements) \cite {Hol01} the
notion of quantum probability measure for a generic variable is
a basic concept. Namely, to each pair $(\hat \rho ,\hat M)$
consisting of the quantum state (the density operator) $\hat \rho $
and the positive operator valued measure $\hat M$, which can be
identifyied with the quantum observable, it is associated
axiomatically (in the way of \cite {MacKey63}) the probability
measure $\mu _{\hat \rho }^{\hat M}$ which determines the readings of
a classical measuring instrument.  Taking into account the enssemble
$\{\mu _{\hat \rho }^{\hat M}\}$ for the fixed measure $\hat M$ and
all states $\hat \rho $, it is possible to reconstruct $\hat M$. In
quantum information theory \cite {Hol98} the states $\hat \rho $ play
a role of quantum probability measures used to encode the
information.  After the states $\hat \rho $ are transmitted through
the quantum channel, one uses $\hat M$ as desicion rules allowing to
decode the data from the channel output. It is worthy to point out
that one can connect a state with a probability distribution density
for arbitrary observable $\hat a$ using the characteristic function
techniques (see \cite {Glauber}). This construction is related to the
construction given by von Neumann in \cite {vonNeu32}, which is used
in quantum information theory \cite {Hol01, Hol98}.

The aim of our work is to establish a new connection of tomographic
probability with
such well studied mathematical object as quantum probability theory
in addition to known
connection with the star-product. Namely, we show that the symplectic
tomogram can be obtained as a partial derivative of the
distribution function of
quantum probability measure. Till now the explicit relations between
the tomographic probability distribution (tomogram) of quantum states
and quantum probability measure concept (for the specific
observable) used in quantum information theory were not known.

The paper is organized as follows: In Section 2 we review the
quantum probability theory. In Section 3 and 4 symplectic tomography
and spin tomography respectively will be discussed. In Section 5
connection of tomogram with quantum probability measure
will be obtained.
In Section 6 new concept of star-product for symplectic
quantum measures will
be introduced. In Section 7 evolution equation for symplectic
quantum
measures will be obtained. Srction 8 is devoted to conclusions and
perspectives.

\section {Basic concepts of quantum probability}

Let $H$ be a separable Hilbert space. A positive linear operator
$\hat \rho $ in $H$ is said to be a state (density operator) if
$Tr\hat \rho =1$.  We denote ${\mathcal L}(H)$, ${\mathcal
L}_+(H)\subset {\mathcal L}(H)$, $\sigma (H)$ and $\hat I$ the sets of
hermitian opeartors (quantum observables), positive operators, states
and the identity operator in $H$ correspondingly.  A map $M$
transmitting each Borel subset ${\Omega}\subset {\mathbb R}$ to a
positive operator $\hat M({\Omega })\in {\mathcal L}_+(H)$ is said to
be a positive operator valued measure (POVM) if $\sum \limits _i \hat
M({\Omega }_i)=\hat I$ for any finite or countable fragmentation
${\mathbb R}=\cup {\Omega }_i$ in the sense of the strong operator
convergence of the series.  The POVM $\hat M$ is said to be
orthogonal if $\hat M({\Omega })^2=\hat M({\Omega })$. By means of
the spectral theorem, given $\hat a\in {\mathcal L}(H)$ there exists
an orthogonal POVM $\hat M$ such that \begin {equation} \hat a=\int
\limits _{\mathbb R}Xd\hat M((-\infty ,X])=\int \limits _{\mathbb
R}X\delta (\hat a -X)dX, \label {ST} \end {equation} where $\delta
(\hat a-X)$ is a density of the measure $\hat M$.  It means that the
measure $\hat M$ is expressed in terms of the Heaviside function $$
\hat M((-\infty
,X])=\theta (\hat a-X)=\frac {1}{2\pi i} \int \limits _{\mathbb
R}\frac {1}{k}e^{-ik(\hat a-X)}dk.
$$
The presentation of the
spectral theorem by means of the operator - valued delta function can
be understood as one can check that matrix elements of last two parts
of (\ref {ST}) in eigenvector basis of the operator $\hat a$ coincide
if we use standard Dirac delta function properties.  In this way, it
is possible to identify observables $\hat a\in {\mathcal L}(H)$ with
orthogonal POVM.  We shall say that $\hat a\in {\mathcal L}(H)$ is
associated with the POVM $\hat M$ appearing in its spectral
representation.  Taking $\hat \rho \in \sigma (H)$ and the POVM
$\hat M$ one can put
\begin {equation}
{\mathcal M}_{\hat \rho
}^{\hat M}({\Omega })=Tr(\hat \rho \hat M({\Omega }))
\label {Q1}
\end {equation}
for any Borel set ${\Omega }\subset {\mathbb R}$.  Then
${\mathcal M}_{\hat \rho }^{\hat M}$ is a classical probability
measure on $\mathbb R$.  Calculating the matrix $\rho (x,x')$ of
the density operator $\hat \rho $ in the basis in which $\hat a$ is
diagonal, we can represent (\ref {Q1}) in the form
\begin {equation}
{\mathcal M}^{\hat M}_{\hat \rho }({\Omega })=\int \limits _{\Omega
}\rho (x,x)dx.
\label {Q1_0}
\end {equation}
If the POVM $\hat M$ is
associated with some observable $\hat a\in {\mathcal L}(H)$, we shall
write ${\mathcal M} _{\hat \rho }^{\hat a}$ as well as ${\mathcal M}
_{\hat \rho }^{\hat M}$.  In quantum probability theory \cite
{Hol01}, it is postulated that ${\mathcal M}_{\hat \rho }^{\hat
M}={\mathcal M}_{\hat \rho }^{\hat a}$ defined by the formula (\ref
{Q1}) gives us a distribution of the readings of a classical
instrument measuring the quantum observable $\hat a$ associated with
POVM $\hat M$ in the state $\hat \rho $. Notice that for the
orthogonal POVM the approach introduced above appeared firstly in
\cite {vonNeu32}. The same probability distribution can be obtained
by means of the characteristic functions techniques \cite {Glauber}.
Alternatively, in quantum information theory \cite {Hol98} $\hat M$
determines certain decision rule allowing to decode the information
containing in the state $\hat \rho $.

For example, consider the position operator $\hat
x\in {\mathcal L}(H)$ acting in the Hilbert space $H=L^2({\mathbb
R})$ by the fromula $(\hat xf)(x)=xf(x),\ f\in H$.  Denote $\chi
_{\Omega }$ the characteristic (indicator) function of the Borel set
${\Omega } \subset {\mathbb R}$ such that $$ \chi _{\Omega }(x)=\left
\{ \begin {array}{c}1,\ x\in {\Omega }\\ 0,\ x\notin {\Omega }\end
{array}\right .  $$ Involve the operator $\hat \chi _{\Omega }$
acting by the formula $(\hat \chi _{\Omega }f)(x)=\chi _{\Omega
}(x)f(x),\ f\in H$.  Then the map $\hat M$ defined for a Borel set
$\Omega $ by the condition $\hat M({\Omega })=\hat \chi _{\Omega }$
is an orthogonal POVM on ${\mathbb R}$.
One can see that $\hat M$ gives us the spectral decomposition of the
operator $\hat x$ such that
\begin {equation}
\hat x=\int \limits _{\mathbb R}X
d\hat \chi _{(-\infty , X]}.
\label {SD}
\end {equation}
Denote $|X><X|$ the density of the orthogonal POVM $\hat \chi$, then
(\ref {SD}) transforms to the following,
$$
\hat x=\int \limits _{\mathbb R}X|X><X|dX,
$$
$$
\int \limits _{\mathbb R}|X><X|dX=\hat I.
$$
In this way, the
probability measure determined by the state $\hat \rho $ can be
written as
\begin {equation}
{\mathcal M} _{\hat \rho }^{\hat x}({\Omega })=Tr(\hat \rho \hat \chi
_{\Omega })=\int \limits _{\Omega }\rho (X,X)dX,
\label {Q2}
\end {equation}
where $\rho (X,X')$ is a matrix of the density operator $\hat \rho $
in the basis consisting of generalized eienvectors $|X><X|$.
For a pure vacuum state $\hat \rho _0=|\psi _0><\psi _0|$ in
the Schrodinger representation
$$
<x|\psi _0>= \frac {1}{(\pi
)^{1/4}}exp(-\frac {x^2}{2})
$$
the formula (\ref {Q2})
yields
\begin {equation}
{\mathcal M} _{\hat \rho _0
}^{\hat x}({\Omega })=\frac {1}{\sqrt {\pi }} \int \limits
_{\Omega }exp(-x^2)dx,
\label {Q3}
\end {equation}
which is a Gaussian
measure with the zero mean and the variance equal to $1/2$.

Let $\hat J_z=\left (\begin {array}{cc}\frac {1}{2}&0\\
0&-\frac {1}{2}\end {array}\right )$ be an operator of spin
projection on the $z$-axis for the particle with the
total spin $J=\frac {1}{2}$. Picking up the Euler angles
$\phi ,\psi ,\theta $ one can define the rotation matrix
by the formula
\begin {equation}
R(\phi ,\psi ,\theta )=
\left (\begin {array}{cc}cos\frac {\theta }{2}
e^{\frac {i(\phi +\psi )}{2}}&isin\frac {\theta }{2}
e^{-\frac {i(\phi -\psi )}{2}}\\
isin\frac {\theta }{2}e^{\frac {i(\phi -\psi )}{2}}&
cos\frac {\theta }{2}e^{-\frac {i(\phi +\psi )}{2}}\end {array}
\right ).
\label {Q4}
\end {equation}
Notice that (\ref {Q4}) determines irreducible representation
of the group $SU(2)$.
It is straihtforward to check that a spectral decomposition
of the operator $\hat a=
R(\phi ,\psi ,\theta )\hat J_zR(\phi ,\psi ,\theta )^{-1}
$
is given by the formula
$$
\hat a=
\left (\begin {array}{cc}\frac {1}{2}cos\theta &-\frac {i}{2}
sin\theta e^{-i\psi }\\
\frac {i}{2}sin\theta e^{i\psi }&-\frac {1}{2}cos\theta \end {array}
\right )=
\frac {1}{2}
\left (\begin {array}{cc}
cos^2\frac {\theta }{2}&-\frac {i}{2}sin\theta e^{-i\psi }\\
\frac {i}{2}sin\theta e^{i\psi }&sin^2\frac {\theta }{2}
\end {array}\right )-
$$
\begin {equation}
\frac {1}{2}
\left (\begin {array}{cc}
sin^2\frac {\theta }{2}&\frac {i}{2}sin\theta e^{-i\psi }\\
-\frac {i}{2}sin\theta e^{i\psi }&cos^2\frac {\theta }{2}
\end {array}\right ).
\label {Q5}
\end {equation}
Using (\ref {Q5}) we obtain for
the probability measure (\ref {Q1})
which is discrete
in the state $\hat \rho =|\psi ><\psi
|$ with $|\psi >=\left (\begin {array}{c} 1\\0\end {array} \right )$
the distribution function as
\begin {equation}
F(x)= {\mathcal M} ^{\hat a}
_{\hat \rho }((-\infty ,x])=\left \{\begin {array}{c}0,\ \ \
\ \ \ \ \ x<-\frac {1}{2}\\ cos^2\frac {\theta }{2},\ -\frac
{1}{2}\leq x<\frac {1}{2}\\ 1,\ \ \ \ \ \ \ \ \ x\geq \frac {1}{2}
\end {array} \right .
\label {Q6}
\end {equation}
The function (\ref {Q6}) gives us the
Bernoulli distribution concentrated in two points $x_1=-\frac {1}{2}$
and $x_2=\frac {1}{2}$ with the probabilities $p=cos^2\frac {\theta
}{2}$ and $1-p=sin^2\frac {\theta }{2}$ correspondingly.

\section {Tomographic representations for continuous variables}

Tomographic probability (tomogram) determining a quantum state is
introduced by relation
\begin {equation}
w(X,\mu ,\nu)=<\delta (X-\mu \hat x-\nu \hat p)>_{\hat \rho },
\label {T1}
\end {equation}
where $\hat x$ and $\hat p$ are position and momentum operators and
the density operator $\hat \rho $ defines the averaging for arbitrary
observable $\hat a$ by
\begin {equation}
<\hat a>_{\hat \rho}=Tr(\hat \rho \hat a).
\label {T2}
\end {equation}
The Dirac $\delta $-function in (\ref {T1}) is defined by its
Fourier decomposition as
\begin {equation}
\delta (\hat a)=\frac {1}{2\pi }\int e^{ik\hat
a}dk.
\label {T3}
\end {equation}
According to \cite {d'Ar96, MMT95, Raymer93} the tomogram (\ref {T1})
is related to the density matrix in position representation $\rho
(y,y')$ by
\begin {equation}
w(X,\mu ,\nu)=\frac {1}{2\pi |\nu |} \int \int \rho
(y,y')e^{\frac {i(y^2-y'^2)\mu } {2\nu }-\frac {iX(y-y')}{\nu }}dydy'
\label {T4}
\end {equation}
and with the Wigner function defined in \cite {Wig32} as
\begin {equation}
W(q,p)=\int \rho (q+\frac {u}{2},q-\frac {u}{2})
e^{-ipu}du
\label {T5}
\end {equation}
via the relation
\begin {equation}
w(X,\mu ,\nu )=\int \int
W(q,p)\delta (X-\mu q -\nu p)\frac
{dqdp}{2\pi }.
\label {T6}
\end {equation}
The tomogram determines the Wigner
function as
\begin {equation}
W(q,p)=\frac {1}{2\pi }\int \int \int
w(X,\mu ,\nu )e^{i(X-\mu q - \nu p)}dXd\mu d\nu .
\label {T7}
\end {equation}
It means that the tomogram can be used to describe the quantum
states completely. In terms of a wave function $\psi (x)$
the tomogram reads for the pure state $\rho =|\psi ><\psi |$
\cite {Mendes}
\begin {equation}
w(X,\mu ,\nu )=\frac {1}{2\pi |\nu |}
|\int \psi (y)e^{\frac {i\mu y^2}{2\nu }-\frac {iXy}{\nu }}dy|^2.
\label {T8}
\end {equation}
Put
$$
\hat \rho _n=|\psi _n><\psi _n|,\ n=0,1,2,\dots ,
$$
\begin {equation}
<X|\psi _n>=\frac {1}{(\pi )^{1/4}}\frac
{1}{\sqrt {2^nn!}}H_n(X)exp(-\frac {X^2}{2})
\label {T8_1}
\end {equation}
are wave functions of the
excited state of an oscillator.
Here $H_n,\ n=0,1,2,\dots $ are the Hermite polynomials.
Then (\ref {T8}) gives us the tomogram for the state $\hat \rho _n$
as
\begin {equation}
w_n(X,\mu ,\nu )=\frac {1}{\sqrt {\pi }}
\frac {1}{2^nn!}
\frac {1}{\sqrt {\mu ^2+\nu ^2}}
H_n^2(\frac {X}{\sqrt {\mu ^2+\nu ^2}})
exp(-\frac {X^2}{\mu ^2+\nu ^2}).
\label {T9}
\end {equation}
Let us involve the wave function
of a coherent state $|\psi _{\alpha }><\psi _{\alpha }|,\
\alpha \in {\mathbb C}$,
$$
<X|\psi _{\alpha }>=\frac {1}{(\pi )^{1/4}}
exp(-\frac {X^2}{2}+\sqrt {2}\alpha X-\frac {\alpha ^2}{2}-
\frac {|\alpha |^2}{2}).
$$
Then the tomogram of the state $|\psi _{\alpha }><\psi _{\alpha }|$
is given by
\begin {equation}
w_{\alpha }(X)=\frac {1}{\sqrt {\pi (\mu ^2+\nu ^2)}}
exp(-\frac {(X-\sqrt {2}Re\alpha \mu -\sqrt {2}Im\alpha \nu )^2}
{\mu ^2+\nu ^2}).
\label {T9_1}
\end {equation}
The most important property of the
tomogram is that it is a standard density of the probability
distribution function on $\mathbb R$, i.e.
\begin {equation}
w(X,\mu ,\nu )\geq 0
\label {T10}
\end {equation}
and
\begin {equation}
\int w(X,\mu ,\nu )dX=1.
\label {T11}
\end {equation}
The physical meaning of the real variable $X$ is that
this variable is equal to position of a particle measured
in rotated and scaled reference frame of the particle
phase space, in which we get
\begin {equation}
X=\mu q+\nu p,\ \mu =e^{\lambda }cos\phi ,\
\nu =e^{-\lambda }sin\phi ,
\label {T12}
\end {equation}
where $\phi $ and $\lambda $ are
an angle of the rotation and a real sqeezing parameter
correspondingly.

\section {Tomography of spin}

Here we review the tomogram of discrete variables (spin).
The tomogram of a spin state is defined by the relation
\begin {equation}
w^{(j)}(m,\theta ,\phi )=
<\delta (m-R^{(j)}(\phi ,\psi,\theta)\hat J_z
R^{(j)}(\phi ,\psi ,\theta )^{-1})>_{\hat \rho},
\label {T13}
\end {equation}
where $\delta $ is the Kronecker operator delta function
which for arbitrary hermitian operator $\hat u$ with
integer eigenvalues is given by the Fourier integral
$$
\delta (\hat u)=\frac {1}{2\pi }\int \limits _0^{2\pi }
e^{i\phi \hat u}d\phi ,
$$
a nonnegative half-integer $j$ is the total spin,
$m=-j,-j+1,\dots ,j$ is the spin projection on $z$-axis,
$\hat J_z$ is a operator of spin projection on the $z$-axis
and $R^{(j)}(\phi ,\psi ,\theta )$ is the matrix of
irreducible representation of the group $SU(2)$ in
standard basis, elements of the group are parametrized by the
Euler angles $\phi ,\psi ,\theta $. The state of spin is
determined by the hermitian nonnegative $2j+1\times 2j+1$ -
density matrix with unit trace. One can show
\cite {OlgaJETP, Dod} that the density operator can be found
from (\ref {T13}). In fact, the formula (\ref {T13}) can be
considered as a linear system of equations determining the density
matrix if the tomogram is a known function.

\section {Quantum probability and tomograms}

By the definition of the orthogonal POVM $\hat M$ given by the
spectral decomposition of the operator $\hat a$, we get $\frac
{d}{dx}\hat M((-\infty ,X])=\delta (X-\hat a)$.  Suppose that
$\hat M$ is
determined by the spectral decomposition of $\mu \hat x+\nu \hat p$.
Comparing (\ref {T1} - \ref {T2}) and (\ref {Q1}) we get
\begin {equation}
w(X,\mu ,\nu )=\frac {d}{dx}{\mathcal M} ^{\mu \hat x+\nu \hat
p}_{\hat \rho } ((-\infty ,X]).
\label {QT1}
\end {equation}
Using the relation
(\ref {QT1}) one can calculate the distribution function of the
measure ${\mathcal M}^{\mu \hat x+\nu \hat p}_{\hat \rho }$ for the
states $\hat \rho _n$ (\ref {T8_1}) in which the tomograms $w_n$ are
given by the formula (\ref {T9}) such that
$$
{\mathcal M} ^{\mu \hat
x+\nu \hat p}_{\hat \rho _n} ((-\infty ,X])=
$$
\begin {equation}
\frac {1}{\sqrt
{\pi (\mu ^2+\nu ^2)}} \frac {1}{2^nn!}\int \limits _{-\infty
}^XH_n^2(\frac {y}{\sqrt {\mu ^2+\nu ^2}}) exp(-\frac
{y^2}{\mu ^2+\nu ^2})dy.
\label {QT1_0}
\end {equation}
Take the function
$$
\phi _{\alpha }(x)=\frac {1}{(\pi )^{1/4}}
exp(-\frac {x^2}{2}+\sqrt {2}\alpha
x- \frac {\alpha ^2}{2})
$$
and decompose it into the series over
the excited wave functions of an oscillator
$\psi _n(x)$ (\ref {T8_1}), then
$$
\phi _{\alpha }(x)=\sum \limits _{n=0}^{+\infty }
\frac {\alpha ^n}{\sqrt {n!}}\psi _n(x).
$$
It follows that
$$
\sum \limits _{n=0}^{+\infty }\sum \limits _{m=0}^{+\infty }
\frac {\alpha ^n}{\sqrt {n!}}\frac {\beta ^m}{\sqrt {m!}}
\int \limits _{-\infty }^X\psi _n(y)
\psi _m(y)dy=\int \limits _{-\infty }^{X}
\phi _{\alpha }(y)\phi _{\beta }(y)dy=
$$
$$
\frac {1}{\sqrt {\pi }}
\int \limits _{-\infty }^{X}exp(-y^2+\sqrt {2}(\alpha +\beta )y-
\frac {\alpha ^2+\beta ^2}{2})dy=
$$
\begin {equation}
\frac {1}{2}e^{\alpha\beta }(1+erf(X-\frac {\alpha +
\beta }{\sqrt {2}})).
\label {QT1_1}
\end {equation}
Using (\ref {QT1_1}) we obtain for the
probability measure (\ref {QT1_0})
$$
{\mathcal M}^{\mu \hat x+\nu \hat p}_{\rho _n}((-\infty ,X])=
$$
$$
\frac {1}{n!}\frac {d^n}{\alpha ^n}\frac {d^n}{\beta ^n}
\{\frac {1}{2}e^{\alpha \beta }(1+
erf(\frac {X}{\sqrt {\mu ^2+\nu ^2}}-\frac {\alpha +\beta }
{\sqrt {2}}))\}|_{\alpha =0,\beta =0}.
$$
In particular, for the
vacuum state $\hat \rho _0$ we have
$$
{\mathcal M}^{\mu \hat x+\nu
\hat p}_{\hat \rho _0}((-\infty ,X])= \frac {1}{2}(1+
erf(\frac {X}{\sqrt {\mu ^2+\nu ^2}})),
$$
which is in accordance
with (\ref {Q3}) if one put $\mu =1,\ \nu =0$.

In this way, using
the tomogram it is possible to obtain the
probability measure ${\mathcal M}^{\mu
\hat x+\nu \hat p}_{\hat \rho }$ without a calculation of the
spectral decomposition for the operator $\mu \hat x+\nu \hat p$.
We shall call ${\mathcal M}^{\mu \hat x+\nu \hat p}_{\hat \rho }$
by a sympectic quantum probability measure.

Analogously, suppose that
the spectral decomposition of the operator
$\hat a=R^{(j)}(\phi ,\psi,\theta)\hat J_zR^{(j)}(\phi ,\psi ,\theta
)^{-1}$ has the following form
$$
\hat a=
\sum \limits _{k=-j}^jk\hat M_k,
$$
where $\hat M_k,\ -j\leq k\leq j,$ form the orthogonal POVM.
It follows from (\ref {T13}) that
$$
w^{(j)}(m,\theta ,\phi )=
\sum \limits _{k=-j}^j\delta _{km}Tr(\hat \rho \hat M_k)=
$$
$$
\sum \limits _{k=-j}^j
\delta _{km}{\mathcal M}
^{\hat a}_{\hat \rho }((k-1,k])=
{\mathcal M} ^{\hat a}_{\hat \rho }((m-1,m]),
$$
where $\delta _{km}$ is the Kronecker symbol, $m=-j,-j+1,\dots ,j$.

\section {Star-product for symplectic quantum measures}

To formulate quantum mechanics using a
map $\hat a\to f_{\hat a}(\vec x)$
between the set of hermitian operators and
the set of functions on appropriate space, one needs
to involve a new multiplication rule for the functions
which should be associative. This multiplication
is said to be a star-product (see \cite {Fronsdal})
and, in particular, it can be involved
for the functions defined by the map $(\vec x=(X,\mu ,\nu ))$
\begin {equation}
f_{\hat a}(\vec x)=Tr(\hat a\delta (X-\mu \hat x-\nu \hat p))
\label {S1}
\end {equation}
in the way of \cite {Man'ko} such that
\begin {equation}
f_{\hat a}*f_{\hat b}(\vec x)=\int K(\vec x_1,\vec x_2,\vec x)
f_{\hat a}(\vec x_1)f_{\hat b}(\vec x_2)d\vec x_1d\vec x_2,
\label {S2}
\end {equation}
where the kernel $K(\vec x_1,\vec x_2,\vec x)$ is written as
$$
K(\vec x_1,\vec x_2,\vec x)=Tr(\hat {\mathcal D}
(\vec x_1)\hat {\mathcal D}(\vec x_2)
\delta (X-\mu \hat x-\nu \hat p))
$$
and $d\vec x=dXd\mu d\nu $.
Here the operator $\hat {\mathcal D}(\vec x)$ is defined by the
formula $$ \hat {\mathcal D}(\vec x)=\frac {1}{2\pi }e^{iX} e^{-i\mu
\hat x-i\nu \hat p}.  $$ The operator $\hat a$ can be reconstructed
from the function $f_{\hat a}(\vec x)$ by the formula \begin
{equation} \hat a=\int \int \int f_{\hat a}(\vec x)\hat {\mathcal
D}(\vec x) dXd\mu d\nu.  \label {S3} \end {equation} In particular,
if $\hat a=\hat \rho\in \sigma (H)$ is a state, then (\ref {S1})
yields a number of the probability distribution densities associated
with $\hat \rho $.

Now consider the map
\begin {equation}
\hat \rho \to {\mathcal M}_{\hat \rho }(\vec x)={\mathcal M}_{\hat
\rho } ^{\mu \hat x+\nu \hat p}((-\infty ,X])
\label {S4}
\end {equation}
from the set of states $\sigma (H)$ to the set of
probability measures on the real line.  The spectral theorem
determines a representation of arbitrary observable $\hat a$ as an
integral over pure states. Hence it is possible to extend the map
(\ref {S4}) from the states $\hat \rho$ to all observables $\hat a$
by means of a linearity.  Taking the spectral decomposition $\mu \hat
x+\nu \hat p=\int Xd\hat M((-\infty ,X])$ we get
\begin {equation}
{\mathcal M}_{\hat a}(\vec x)=Tr(\hat
a\hat M((-\infty ,X]))).
\label {S5}
\end {equation}
Here for an arbitrary observable $\hat
a$ we obtain the family
${\mathcal M}_{\hat a}(\vec x)$ determining
a non-positive Borel measure on the real line
for any fixed $\mu $ and $\nu $.
We shall call them by symplectic quantum measures.
Comparing (\ref {S3}),(\ref {QT1}) and
(\ref {S5}) we obtain
\begin {equation}
\frac {d}{dX}{\mathcal M}_{\hat a}(\vec x)=f_{\hat a}(\vec x)
\label {S6}
\end {equation}
and the operator
$\hat a$ can be reconstructed from the family of
symplectic quantum measures
${\mathcal M}_{\hat a}$ such that
\begin {equation}
\hat a=\int \int \int
\hat {\mathcal D}(\vec x)d{\mathcal M}_{\hat a}(X)d\mu d\nu.
\label {S7}
\end {equation}
Here $d{\mathcal M}_{\hat a}(X)$ denotes the measure
on the real line determined by ${\mathcal M}_{\hat a}(\vec x)$
for the fixed $\mu $ and $\nu $.
In the formula (\ref {S7}) and below
we claim that, at first,
the integration is done over $d{\mathcal M}_{\hat a}(X)$
for the fixed $\mu ,\nu $
and then we integrate over $d\mu $ and $d\nu $.
The formula (\ref {S7}) allows to define an associative
multiplication for the symplectic quantum
measures ${\mathcal M}_{\hat a}$
which can be derived from (\ref {S2}),
\begin {equation}
{\mathcal M}_{\hat a}*{\mathcal M}_{\hat b}(\vec x)=\int \tilde
K(\vec x_1,\vec x_2,\vec x)
d{\mathcal M}_{\hat a}(X_1)d\mu _1d\nu _1d{\mathcal M}_{\hat b}(X_2)
d\mu _2d\nu _2,
\label {S8}
\end {equation}
where the kernel
$\tilde K(\vec x_1,\vec x_2,\vec x)$ defined by the formula
$$
\tilde
K(\vec x_1,\vec x_2,\vec x)=Tr(\hat {\mathcal D}(\vec x_1)\hat
{\mathcal D}(\vec x_2)\hat M((-\infty ,X])), $$ $\vec x_1=(X_1,\mu
_1,\nu _1),\ \vec x_2=(X_2,\mu _2,\nu _2)$.  By a construction the
multiplication (\ref {S8}) is in accordance with the obvious
multiplication on the set of observables, such that \begin {equation}
{\mathcal M}_{\hat a\hat b}(\vec
x)={\mathcal M}_{\hat a}* {\mathcal M}_{\hat b}(\vec x).
\label {S9}
\end {equation}
It follows
from (\ref {S9}) that the symplectic
probability measures ${\mathcal M}_{\hat
\rho }$ associated with the pure states $\hat \rho =|\psi ><\psi |$
are idempotents with respect to the multiplication $*$, i.e.
$$
{\mathcal M}_{|\psi ><\psi |}*{\mathcal M}_{|\psi ><\psi |}=
{\mathcal M}_{|\psi ><\psi |}.
$$

\section {Evolution for symplectic quantum measures}

It was shown in \cite {MMT96} that if the quantum state
satisfies the time-evolution equation
\begin {equation}
\partial _t\hat \rho =-i[\hat H,\hat \rho ],
\label {E1}
\end {equation}
then for systems with Hamiltonian
of the form $\hat H=\frac {\hat p^2}{2}+V(\hat q)$
the evolution of the quantum tomogram (\ref {T1}) by virtue of
(\ref {E1}) is given as
$$
\frac {\partial w}{\partial t}-\mu \frac {\partial w}{\partial \nu }-
$$
\begin {equation}
-i(V(-
(\frac {\partial }{\partial X})^{-1}\frac {\partial }{\partial \mu }
-\frac {i\nu }{2}
\frac {\partial }{\partial X})-
V(-(\frac {\partial }{\partial X})^{-1}\frac {\partial }
{\partial \mu}+\frac {i\nu }{2} \frac {\partial }{\partial X}))w=0.
\label {E2}
\end {equation}
It is straightforward to check that the equation (\ref {E2})
determines the evolution of the functions (\ref {S1}) and,
therefore, by means of the connection (\ref {S6}),
of the distribution functions ${\mathcal M}_{\hat a}(X)$
of symplectic quantum measures (\ref {S5}).
Integrating (\ref {E2}) over $X$ we
obtain the following form of evolution,
$$
\frac {\partial {\mathcal
M}_{\hat a}(X)}{\partial t}-
\mu \frac {\partial {\mathcal M}_{\hat a}(X)
}{\partial \nu }-
$$
\begin {equation}
-i(V(-
(\frac {\partial }{\partial X})^{-1}\frac {\partial }{\partial \mu }
-\frac {i\nu }{2}
\frac {\partial }{\partial X})-
V(-(\frac {\partial }{\partial X})^{-1}\frac {\partial }
{\partial \mu}+\frac {i\nu }{2} \frac {\partial }{\partial X}))
{\mathcal M}_{\hat a}(X)=0.
\label {E3}
\end {equation}
Note that we have derived the linear evolution equation
valid only for symplectic quantum measures.
The general equation for continuous nondemolition
measurement introduced in \cite {Bel90'1, Bel90'2} is
nonlinear.

\section {Conclusion}

To conclude we summarise the main results of the paper.
We established relation (see Equ. (\ref {QT1})) of symplectic
tomograms to distribution functions of
symplectic
quantum probability measures, which gives a possibility
to use results of the quantum probability theory to
study properties of quantum tomograms.
On the base of the established relation the
evolution equation for the symplectic (non-positive in general)
quantum measures is obtained (see Equ. (\ref {E3})). The
star product of symplectic quantum measures is introduced
(see Equ. (\ref {S8})).
The approach is
illustrated by example of several different states of
harmonic oscillator.
In addition, the quantum probability
measures and its connection with tomogram are constructed
for such discrete observable as spin.

\end {document}